\documentclass[%
 aip,
rsi,%
 amsmath,amssymb,
preprint,%
 reprint,%
 groupedaddress,
]{revtex4-1}

\usepackage{graphicx}
\usepackage{dcolumn}
\usepackage{bm}
\usepackage{caption}
\usepackage{hyperref}
\usepackage[export]{adjustbox}
\hypersetup{
    colorlinks=true,
    linkcolor=magenta,
    urlcolor=blue,
    citecolor=red,
}
\usepackage{enumitem}
\usepackage{xcolor}
\usepackage{siunitx}
\usepackage{subcaption}
\usepackage{relsize}

\usepackage[draft]{todonotes}   

\begin{document}

\preprint{APS/123-QED}

\title{Microcontroller based scanning transfer cavity lock for long-term laser frequency stabilization}

\author{S. Subhankar}
  \email{sarthaks@umd.edu.}
\author{A. Restelli}%
\author{Y. Wang}
\author{S. L. Rolston}
\author{J. V. Porto}
\affiliation{
Joint Quantum Institute, National Institute of Standards and Technology 
and the University of Maryland, College Park, Maryland 20742 USA}

\date{\today}
\begin{abstract}
We present a compact all-digital implementation of a scanning transfer cavity lock (STCL) for long-term laser frequency stabilization. An interrupt-driven state machine is employed to realize the STCL, with the capability to correct for frequency drifts in the slave laser frequency due to measured changes in the lab environmental conditions. We demonstrate an accuracy of  0.9 MHz for master laser and slave laser wavelengths of $556$~nm and $798$~nm as an example. The slave laser is also demonstrated to dynamically scan over a wide frequency range while retaining its lock, allowing us to accurately interrogate atomic transitions.   
\end{abstract}

\pacs{Valid PACS appear here}
\maketitle
\section{Introduction}
Many applications require stabilizing the frequency of a laser, and various methods have been developed to lock a laser frequency to a desired value. One of the simplest locking techniques is the scanning transfer cavity lock (STCL)~\cite{Burke2005,Wang2013,Seymour-Smith2010, Rossi2002,Zhao1998} in which
the stability of a master laser frequency (for example locked to an atomic transition) is transferred to a scanned Fabry-Perot cavity, which plays the role of a frequency discriminator, and the slave laser frequency is then stabilized to the cavity. In addition to its simplicity, STCL has a wide capture range and the difference in wavelength between the slave and master lasers can be multiple nanometers. Scanned Fabry-Perot cavities are common in atomic physics labs, and the optical hardware for SCTL is typically readily available.
\begin{figure}
	\centering
\includegraphics[width=8.5cm]{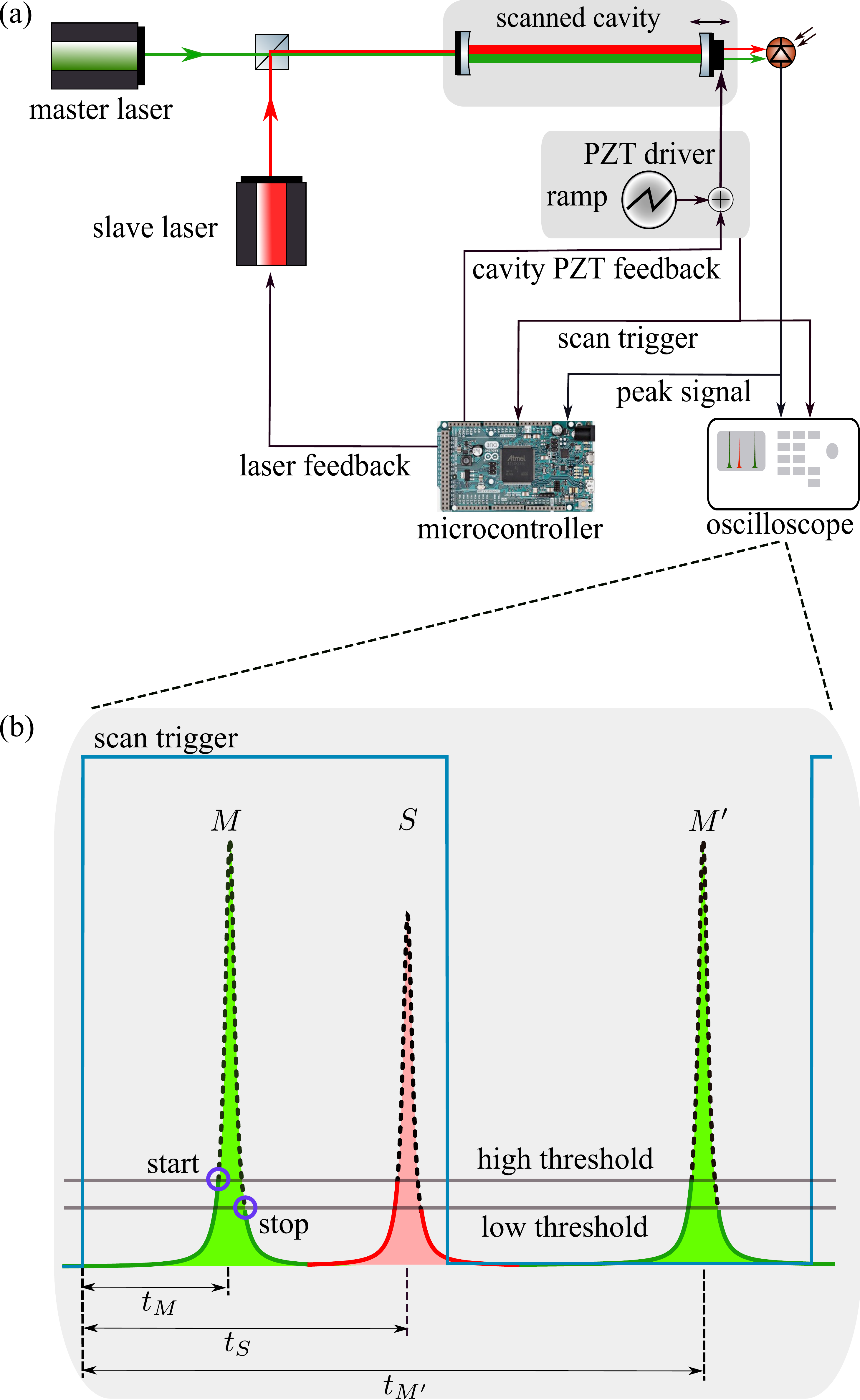}
\caption{\textbf{(a)} Schematic of the STCL hardware.  \textbf{(b)} Schematic of the STCL peak finding algorithm. The dashed segments of the peaks represent the relevant peak data that are acquired and processed by the $\mu$C to determine the arrival times of the peaks, $t_M$, $t_S$ and $t_{M^{\prime}}$. All the arrival times are measured with respect to the rising edge of the scan trigger. The circles pinpoint the (start timestamp, high threshold) and (stop timestamp, low threshold) time-voltage level coordinates between which the peak data of interest (dashed) is acquired. }
\label{fig:scanpic}
\end{figure}

In this paper, we present an all-digital and cost-effective approach for implementing the STCL. The signal acquisition and processing, detailed in Sec.~\ref{statemachine}, is done digitally in a low-cost Arduino Due development board~\cite{discliamer} mounted on a custom shield. This is in contrast to the implementations that use analog circuitry for peak detection~\cite{Burke2005,Wang2013,Seymour-Smith2010} and a dedicated PC for signal acquisition and/or processing~\cite{Seymour-Smith2010,Zhao1998,Rossi2002}. We also investigate the effect of the environment on the slave laser frequency~\cite{Uetake2009, Matsubara2005} and present a method that can compensate for this effect to within an accuracy of 0.9 MHz for master  and slave laser wavelengths of $556$ nm and $798$ nm respectively. We monitor the environment using a low-cost commercially available BME280 sensor breakout board~\cite{discliamer} that measures the temperature, pressure and humidity with an accuracy and precision sufficient for the measurements we present in Sec.~\ref{envt}. This sensor can be integrated to the current hardware design with relevant updates to the current software protocol to provide a compact all-digital STCL with long-term laser frequency stability and accuracy. The particulars of our STCL implementation are detailed in Sec.~\ref{setup}. The Github link for the project is: \href{https://github.com/JQIamo/Scanning-Transfer-Cavity-Lock}{https://github.com/JQIamo/Scanning-Transfer-Cavity-Lock}.   

\section{Scanning Transfer Cavity Lock Technique}
The transmission resonances of light through a Fabry-Perot cavity relate the frequency of the light to the length of the cavity. For a confocal cavity, the resonance frequencies are given by
\begin{equation}
\nu=\frac{N c}{4 n d}
\label{cavity}
\end{equation}
where $\nu$ is the laser frequency, $N$ is the longitudinal mode number of the resonance, $c$ is the speed of light, $n$ is the refractive index of the medium inside the cavity, and $d$ is the length of the cavity. For a fixed cavity length, the transmission peaks are spaced by the free spectral range (FSR): $\Delta_{\textrm{\textrm{FSR}}} = c/4n d$. 
In our implementation, the length of the cavity is scanned with an amplitude large enough such that the resonant frequency is scanned over a range slightly larger than its FSR. The average cavity length is adjusted to provide three peaks arranged in a Master-Slave-Master ($M-S-M^\prime$) configuration as shown in Fig.~\ref{fig:scanpic}{\color{magenta}b}. For a linear scan of $d$ with speed $\alpha$, the arrival time of the peak, $t_i$ is given by
$$ t_i=(d_i-d_0)/\alpha ,$$
where $i=M$, $S$ or $M^\prime$, $d_i$ is the resonant cavity length and $d_0$ is the (arbitrary) cavity length at $t=0$. Using Eq.~\ref{cavity}, we relate the frequencies of the lasers to the arrival times of the peaks and  provide signals that can be used to stabilize the average cavity length and the frequency of the slave laser. Drifts in the average cavity length are measured by the position of the first peak of the master laser, $t_M$, which is used to stabilize $d_0$. With the average cavity length locked, $t_S$ is then used to stabilize the desired slave laser frequency $\nu_S$ to the cavity. In order to remove dependence on the ramp speed $\alpha$, we use the second peak position of the master laser, $t_{M^\prime}$, forming the ratio
\begin{equation}
r=\frac{t_{M}-t_{S} }{t_{M^\prime}-t_{M}}=\frac{d_{M}-d_S}{d_{M^\prime}-d_M}=N_M-N_S\frac{ n_M \nu_M}{n_S\nu_S}.
\label{ratio}
\end{equation}
where $n_i$ is the refractive index of air for light at frequency $\nu_i$ and $i=M,S$.
The slave mode number $N_S$ is the largest integer smaller than $N_M (n_S \nu_S)/(n_M \nu_M)$, so that $0<r<1$. If we define a reference frequency $\nu_{S0} = (\nu_M n_M N_S)/(n_SN_M)$, then the slave frequency $\delta \nu_S = \nu_{S}-\nu_{S0}$ is given by
\begin{equation}
\delta \nu_S=\nu_M \frac{N_S}{N_M}\frac{n_M}{n_{S}}\left(\frac{r/N_M}{1-r/N_M} \right) \simeq \frac{N_S}{N_M} \Delta_{\textrm{FSR}}\ r,
\label{slavefreq}
\end{equation}
up to order $r/N_M\ll 1$ and $n_M,n_S\simeq1$. Deviations of $r$ from a chosen lock point $r_0$ generate the error signal that can be used to feedback to the slave laser.

\section{Software Implementation} \label{statemachine}

In order to lock both the cavity to the master laser and the slave laser to the cavity via $t_M$ and $r$, the arrival times ($t_M,t_S,t_{M^\prime}$) of the peaks need to be determined during each scan of the cavity length, the respective error signals calculated, and feedback performed via the changes in control voltages to the piezoelectric transducers (PZT) that sets the DC offset for the cavity spacing $d_0$ and the slave laser frequency. We use an interrupt-driven~\cite{White2011,Yiu2013} state machine to achieve these tasks.

\subsection{State machine}

\begin{figure}[h]
	\centering
	\includegraphics[width=8.5cm]{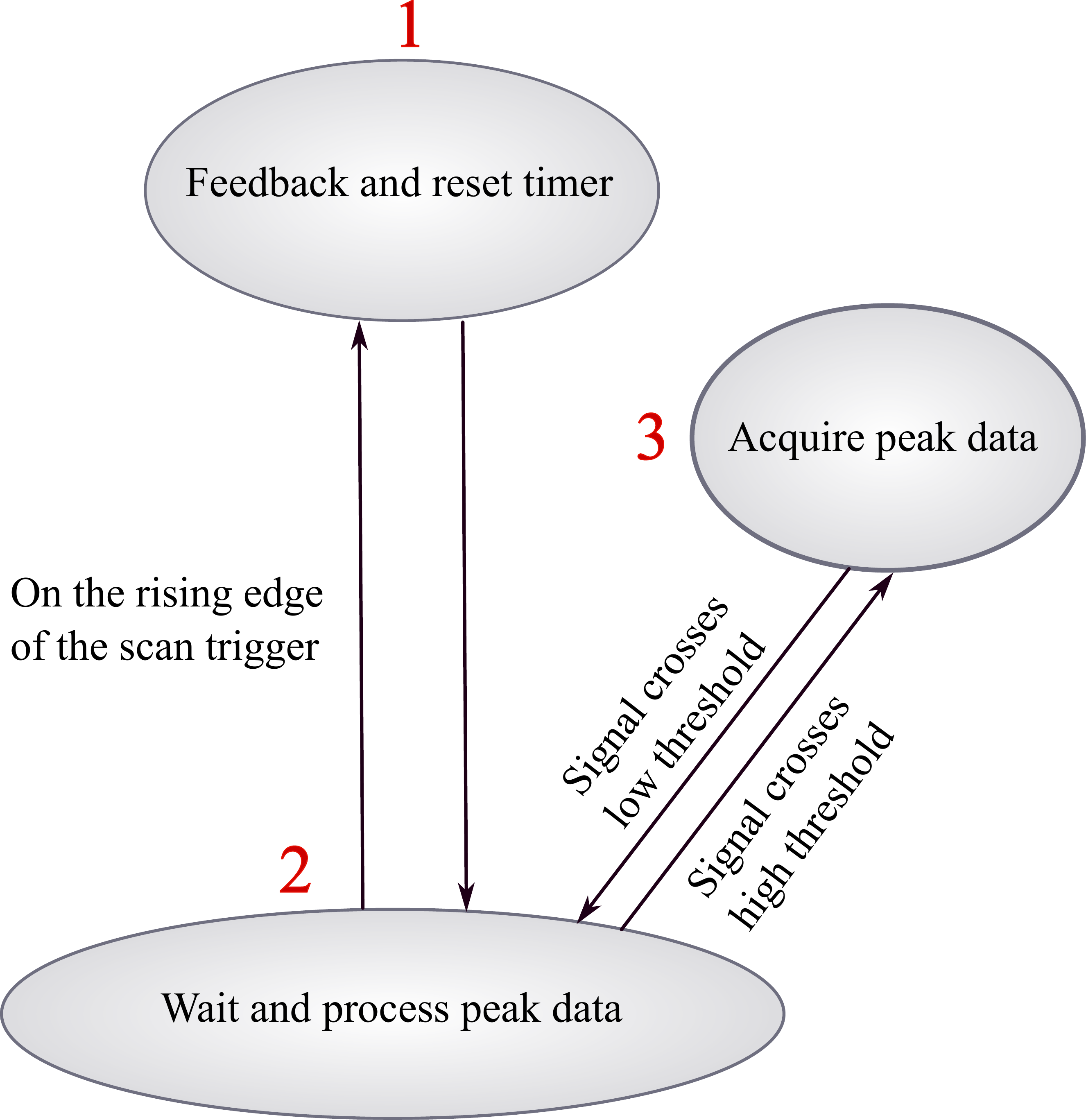}
	\caption{State machine schematic. The arrows indicate the transitions between states and their directions upon meeting the stated requirements. }
	\label{fig:statemachine}
\end{figure}

 The flow structure of the interrupt-driven state machine is shown in Fig.~\ref{fig:statemachine}. It is designed to respond to a series of interrupts generated by hardware (peripherals) or software (software triggered interrupts) which changes the control flow of execution in the program~\cite{White2011,Yiu2013}.

The functionalities of each state and the conditions for transitions between states is described below:
\begin{enumerate}[label=\color{red}\textbf{\arabic*}:]
\item State {\color{red}1} is the feedback and reset timer state. It is triggered by an interrupt on the rising edge of the scan trigger (Fig.~\ref{fig:scanpic}{\color{magenta}b}) provided by the cavity driver. It updates the control voltages to the slave laser and the Fabry-Perot cavity that were calculated in the previous cycle and the system timer is reset. All timestamps are referred to the rising edge of the scan trigger (Fig.~\ref{fig:scanpic}{\color{magenta}b}). Upon completion, it returns to state {\color{red}2}.

\item State {\color{red}2} is the data processing and wait state.
If returning from state {\color{red}1}, it waits until a peak signal is ready to be sampled. When the signal exceeds the high threshold of the programmed comparison window (Fig.~\ref{fig:scanpic}{\color{magenta}b}) the Analog to Digital Converter (ADC) asserts an interrupt to change the state to state {\color{red}3}. If returning from state {\color{red}3}, it processes the acquired data and waits.
Processing involves finding the arrival time of the peak ($t_M$ / $t_S$ / $t_{M^{\prime}}$) and calculating the new control signals for the cavity and the slave laser feedback. Upon completion, it waits.

\item  State {\color{red}3} is the data acquisition state. The start timestamp is saved and the data acquisition is initiated. When the peak data crosses the low threshold, the ADC asserts an interrupt which saves the stop timestamp and terminates the data acquisition and returns to state {\color{red}2}. The sampled data is depicted by the dashed segments in Fig.~\ref{fig:scanpic}{\color{magenta}b}.   

\end{enumerate}

\subsection{Data acquisition}

 The transmission peaks are sampled by a 12-bit ADC native to the SAM3x8E $\mu$C~\cite{discliamer} at 1 Megasamples per second. In order to optimize RAM usage and reduce the need for data filtering of irrelevant data, we only save sampled data near the peak via signal threshold based interrupts in our state machine. We use the direct memory access functionality to rapidly transfer the data of interest directly into a buffer in the $\mu$C RAM without any processor intervention. We found that a comparison-window based interrupt (high threshold-low threshold as shown in Fig.~\ref{fig:scanpic}{\color{magenta}b})\cite{Yiu2013,SAM3x} is superior to a single-valued level-triggered interrupt, since fluctuations in the peak signal near the threshold spuriously triggered interrupts. This is resolved by setting a sufficient difference between high threshold and low threshold. We typically sample 125 points for each peak.

\subsection{Peak detection algorithm}
\label{peakdetect}
The position of the maximum of the transmission peak is given by the zero-crossing of the 1st derivative determined by a 5-point digital Savitzky-Golay (SG) filter~\cite{Howard2004,Schafer2011}. This filter is an efficient method to smooth the acquired data without significantly distorting the signal while improving the Signal-to-Noise ratio. It has the following form
\begin{equation}
Y^\prime_j=\frac{-2y_{j-2}-y_{j-1}+y_{j+1}+2y_{j+2}}{10}
\label{filter}
\end{equation}
where $y_j=$ value of the buffer at index $j$ and $Y^\prime_j=$ derivative at buffer position $j$. This filter can be efficiently implemented using shift operators in the program. Using the zero-crossing timestamps to tag the peaks makes the STCL robust to laser power fluctuations. Upon finding a zero-crossing timestamp ($t_M$, $t_S$ or $t_{M^\prime}$) the state machine calculates the error signals and control voltages for the servo loop.

\subsection{Servo loop}
\label{servo}

The servo loop feeds back on the slave laser and to the cavity PZT to stabilize the cavity length. The zero-crossing timestamps $t_M,t_S$ and $t_{M^\prime}$ are used to compute the error signal for the cavity and slave laser, which are $t_{M,\textrm{ lock}}-t_M$ and $r_{\textrm{lock}}-\Delta t_{MS}/\Delta t_{MM^\prime}$ respectively. The control signal ($u(t_y)$) at discrete time $t_y$ is given by the discrete PI filter~\cite{Astrom2006}
\begin{equation}
u(t_y)=u(t_{y-1})+K_\textrm{P} (e(t_y)-e(t_{y-1}))+K_\textrm{I} e(t_y)\Delta t 
\label{pilock}
\end{equation}
where $e(t_y)$ is the error signal at $t_y$, $K_\textrm{P}$ is the proportional gain, $K_\textrm{I}$ is the integral gain, $\Delta t$ is the time it takes to scan the cavity, and $t_y=y\Delta t$ where $y$ is an integer. 
Updates are performed at the rising edge of the next scan trigger using two 12-bit DACs native to the SAM3x8E $\mu$C. Anti-windup is implemented by not updating the value of $u(t_y)$ if $u(t_y)-u(t_{y-1})$ causes the DAC output to fall outside an adjustable range of voltages.
\begin{figure}
\centering
\includegraphics[height=2.2in]{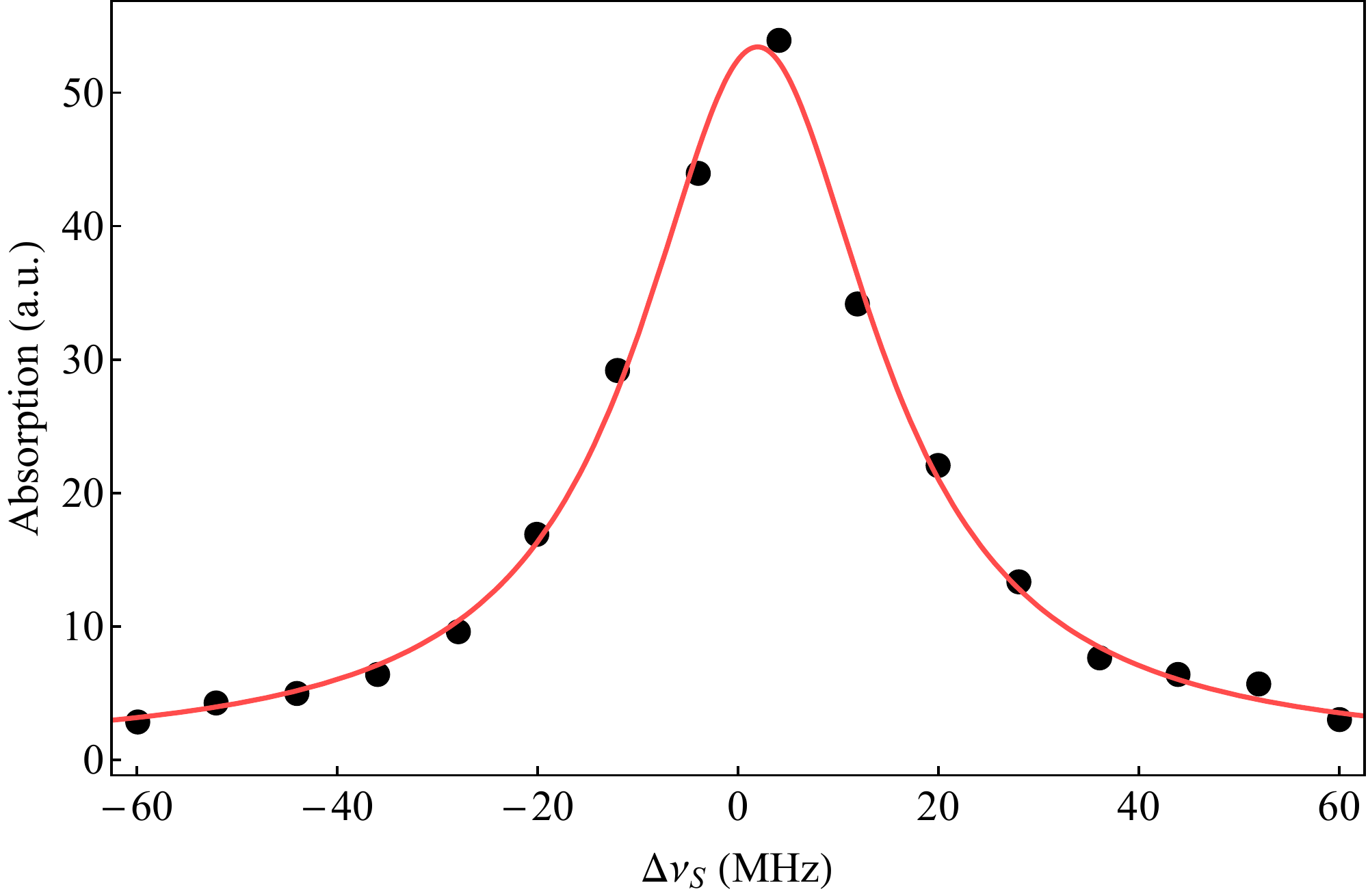}
\caption{Absorption spectrum of the $^1S_0\leftrightarrow{}^1P_1$ transition in Yb as the slave laser frequency is stepped through a chosen frequency range in every experimental realization. The Lorentzian fit to the distribution gives a linewidth $\Gamma=2\pi\times(28.74\pm1.16)$ MHz which is close to the natural linewidth of $2\pi\times28$ MHz.}
\label{fig:scan}
\end{figure}

\section{Experimental setup}
\label{setup}
 In our implementation, we use a Thorlabs scanning confocal Fabry-Perot Cavity, SA200-5B~\cite{discliamer}, that has an FSR of 1.5 GHz. The cavity PZT is scanned in a sawtooth fashion with a period of 100 Hz and with an amplitude that scans the resonant frequency of the cavity by 1.2 FSR. The cavity is neither temperature controlled nor sealed or evacuated.
 
 For the master, we demonstrate the lock with two different lasers, one at 780~nm and one at 556~nm, and for the slave we use a laser at 798~nm. The $\lambda_M=780$ nm master laser is stabilized to a saturated absorption feature on the $^{85}$Rb $5^2S_{1/2}|F=3\rangle \leftrightarrow 5^2P_{3/2}|F^\prime=3-4\rangle$ crossover signal with a linewidth of 1 MHz and the other $\lambda_M=556$ nm master laser is stabilized to the $^1S_0|F=1/2\rangle\leftrightarrow{}^3P_1|F=3/2\rangle$ transition in $^{171}$Yb with a linewidth of 1 MHz. Once the cavity is stabilized to either of the two master lasers, the slave laser is locked to the cavity, using the ratio $r$. The locked slave laser at $\lambda_S=798$ nm is the seed input for a Toptica TA-DL SHG pro laser system~\cite{discliamer} that generates frequency-doubled light at 399 nm that we use to interrogate the $^1S_0\leftrightarrow{}^1P_1$ transition in Yb.
 
\section{Performance}
\subsection{Lock bandwidth}
In our implementation, the bandwidth of the lock is limited by the frequency of the cavity PZT scan to 100 Hz. The state machine on the Arduino Due development board can acquire peak data, process it, and update the feedback output voltages for each ramp of the cavity at a maximum rate of 2 kHz, which is much faster than the 100~Hz cavity PZT scan. Pound-Drever-Hall locking~\cite{Drever1983} to a cavity has a much higher bandwidth allowing one to narrow the linewidth of a laser, but is more involved as it requires modulating the laser frequency and demodulating the photodiode signal. The STCL is simpler to implement and is intended for stabilization against long-term laser frequency drifts, and does not narrow the slave laser (the slave laser in our experiment has an intrinsic short-term laser linewidth of $\sim100$~kHz). The ultimate limit on the bandwidth of the STCL will likely be determined by the speed of the cavity PZT scan.  
\subsection{Dynamic setpoint change} 
The STCL allows us to scan the frequency of the slave laser by changing $r$. We translate a change in $r$, $\Delta{r}$ , to a corresponding change in frequency of the slave laser, $\Delta{\nu_S}$, through a scale factor, $\beta$. The value of $\beta$ can be calibrated using atomic transitions or calculated via first principle as follows: 
\begin{align}
&r+\Delta r=N_M-N_S\frac{\nu_M}{(\nu_S+\Delta\nu_S)}\\
\implies& \Delta r\simeq\frac{N_S\nu_M}{\nu^2_S}\Delta{\nu_S}=\frac{4d\lambda_S}{c\lambda_M}\Delta{\nu_S}=\beta^{-1} \Delta{\nu_S}
\label{changefactor}
\end{align}
where $n_M,n_S\simeq1$. With $d=50$ mm, $\lambda_M=780 $ nm, and $\lambda_S=798$ nm, $\beta=1.465$ GHz. In our implementation, it takes 40 ms for the STCL to lock after a sudden jump in its slave laser setpoint $r$. In principle, since the slave laser response time is much faster than the 10 ms cavity sweep time, by feed-forwarding on the slave laser control voltage it should be possible to change the slave laser frequency in one cavity sweep. 
\paragraph*{}
In Fig.~\ref{fig:scan}, we show the absorption spectrum of the  $^1S_0\leftrightarrow{}^1P_1$ transition in $^{171}$Yb  obtained by scanning the frequency of the slave laser using the STCL. The linewidth extracted from the fit matches well with the natural linewidth of the transition, indicating that the magnitude of $\beta$ has been accurately determined.
\begin{figure}
	\centering
	\label{edlentheory}
	\begin{subfigure}[t]{0.5\textwidth} 
		\subcaption{}
		\includegraphics[height=2.35in]{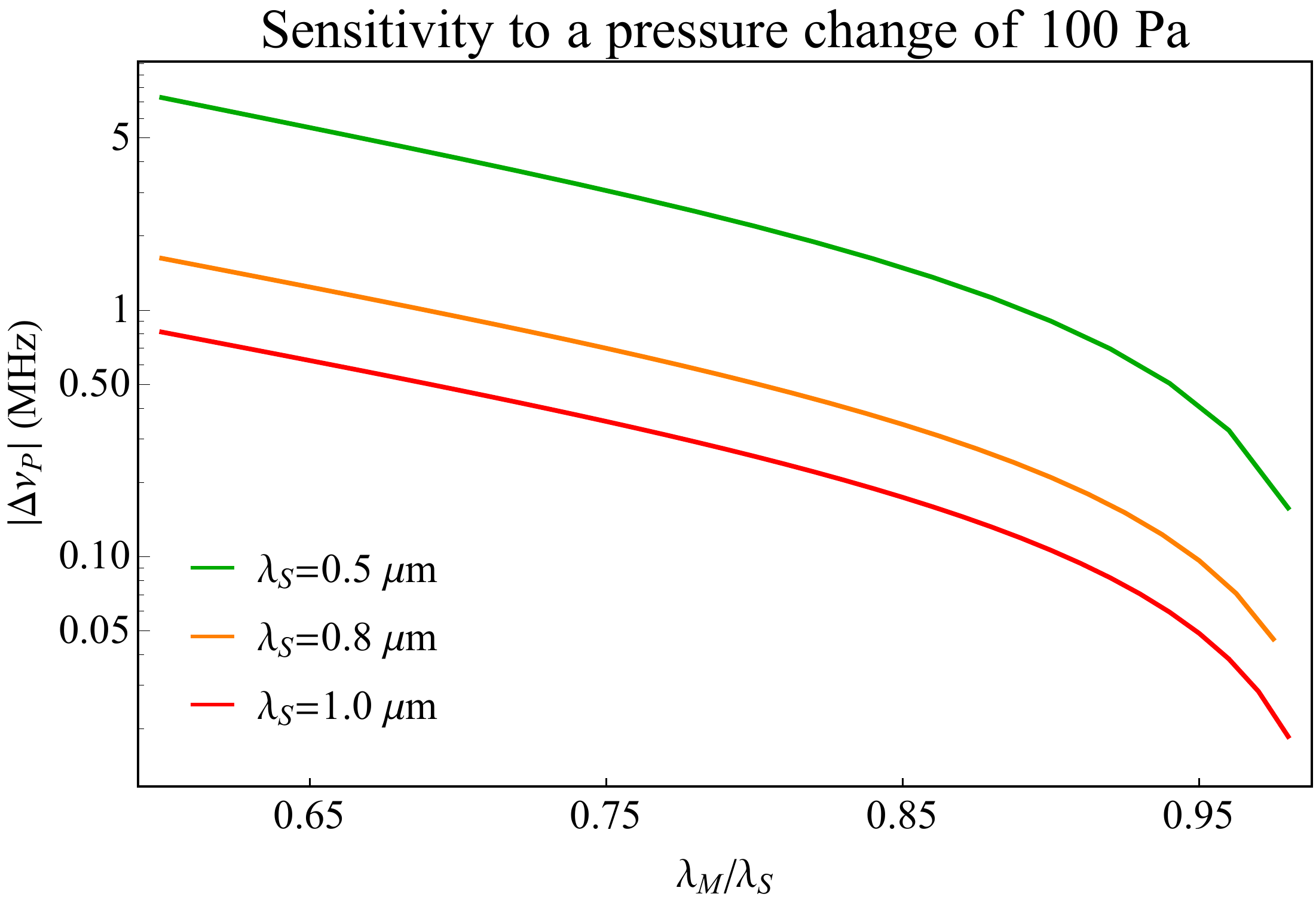}
		\centering
		\label{Theory600}
		
	\end{subfigure}%
	
	\begin{subfigure}[t]{0.5\textwidth} 
		\subcaption{}
		\includegraphics[height=2.35in]{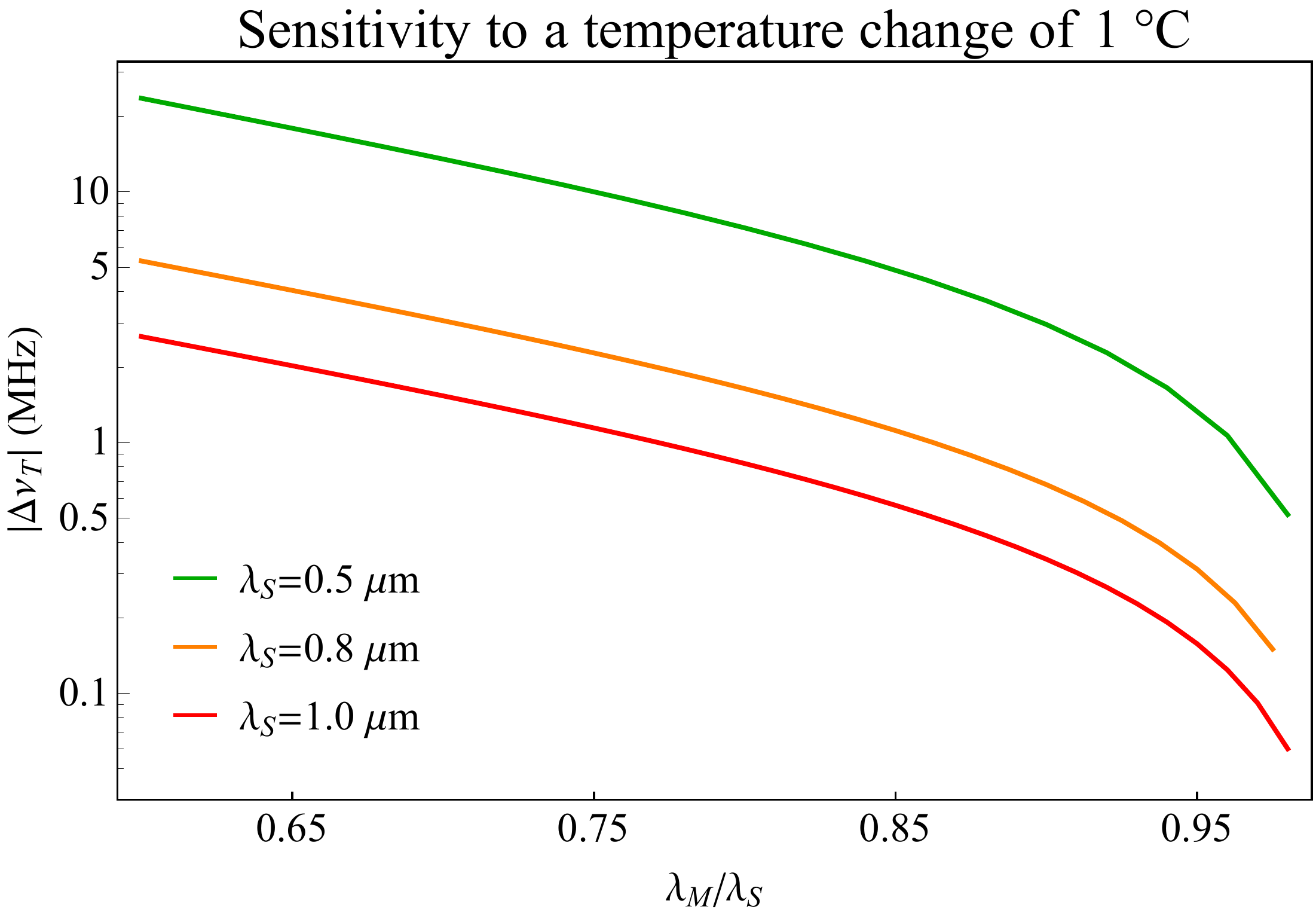}
		\centering
		\label{Theory800}
		
	\end{subfigure}
	
	\begin{subfigure}[t]{0.5\textwidth} 
		\centering
		\subcaption{}
		\includegraphics[height=2.35in]{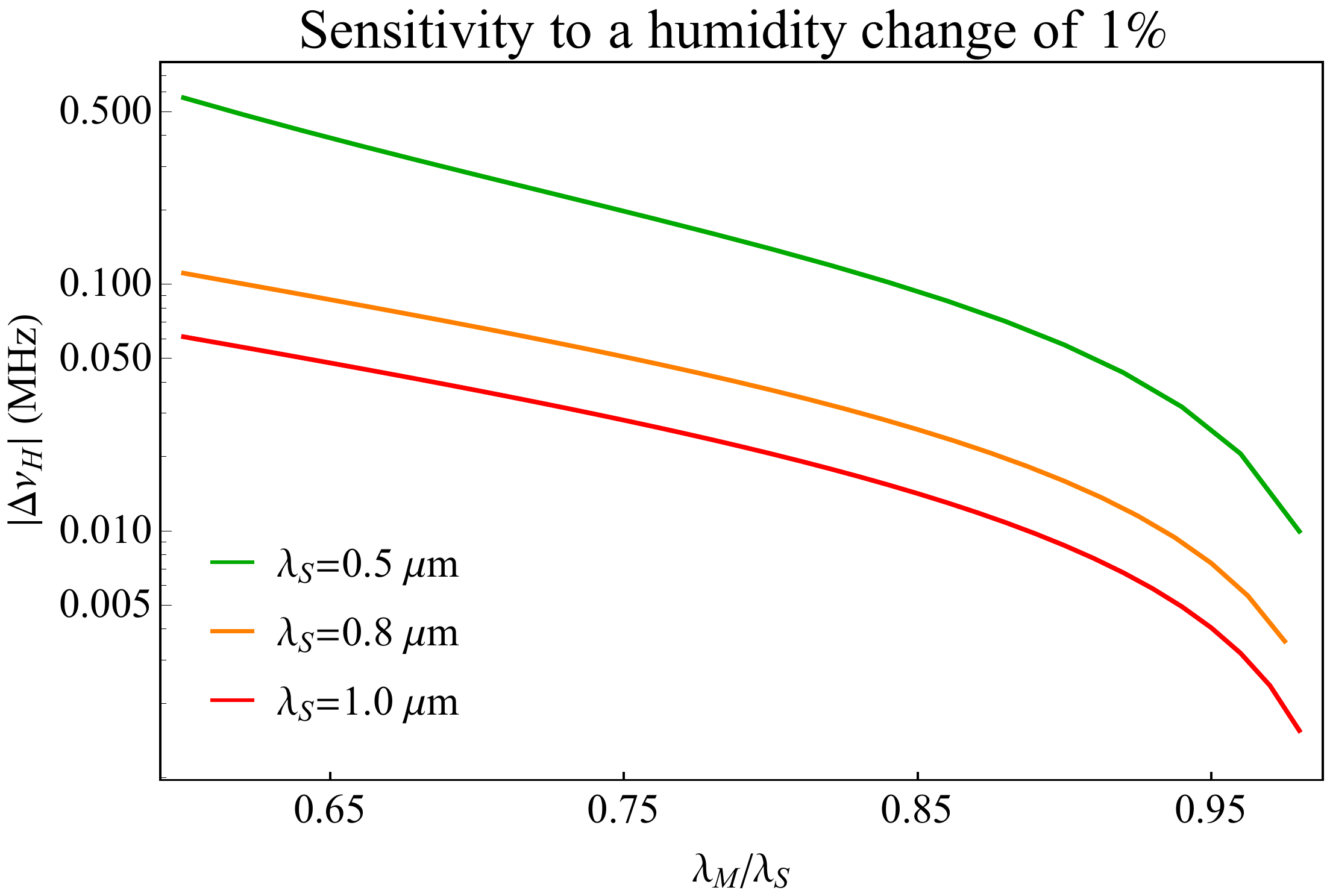}
		\label{Theory1000}
	\end{subfigure}%

	\caption{Calculations of the sensitivity of the slave laser frequency, $\Delta\nu_{X_i}$ , to changes in environmental parameters $\Delta X_i$ ($\Delta$T = 1~$^{\circ}$C, $\Delta$P = 100~Pa, or $\Delta$H = 1~\%) as a function of master and slave laser wavelengths. The absolute values of the sensitivity coefficients monotonically decrease as  $\lambda_M/\lambda_S$ approaches 1, implying that the closer the wavelengths $\lambda_M$ and $\lambda_S$ are to each other, the less sensitive is the accuracy of the slave laser frequency to variations in environmental parameters.}
\end{figure}

\begin{figure*}
	\centering
	\begin{subfigure}[t]{0.5\textwidth} 
		\subcaption{}
		\includegraphics[height=2.4in]{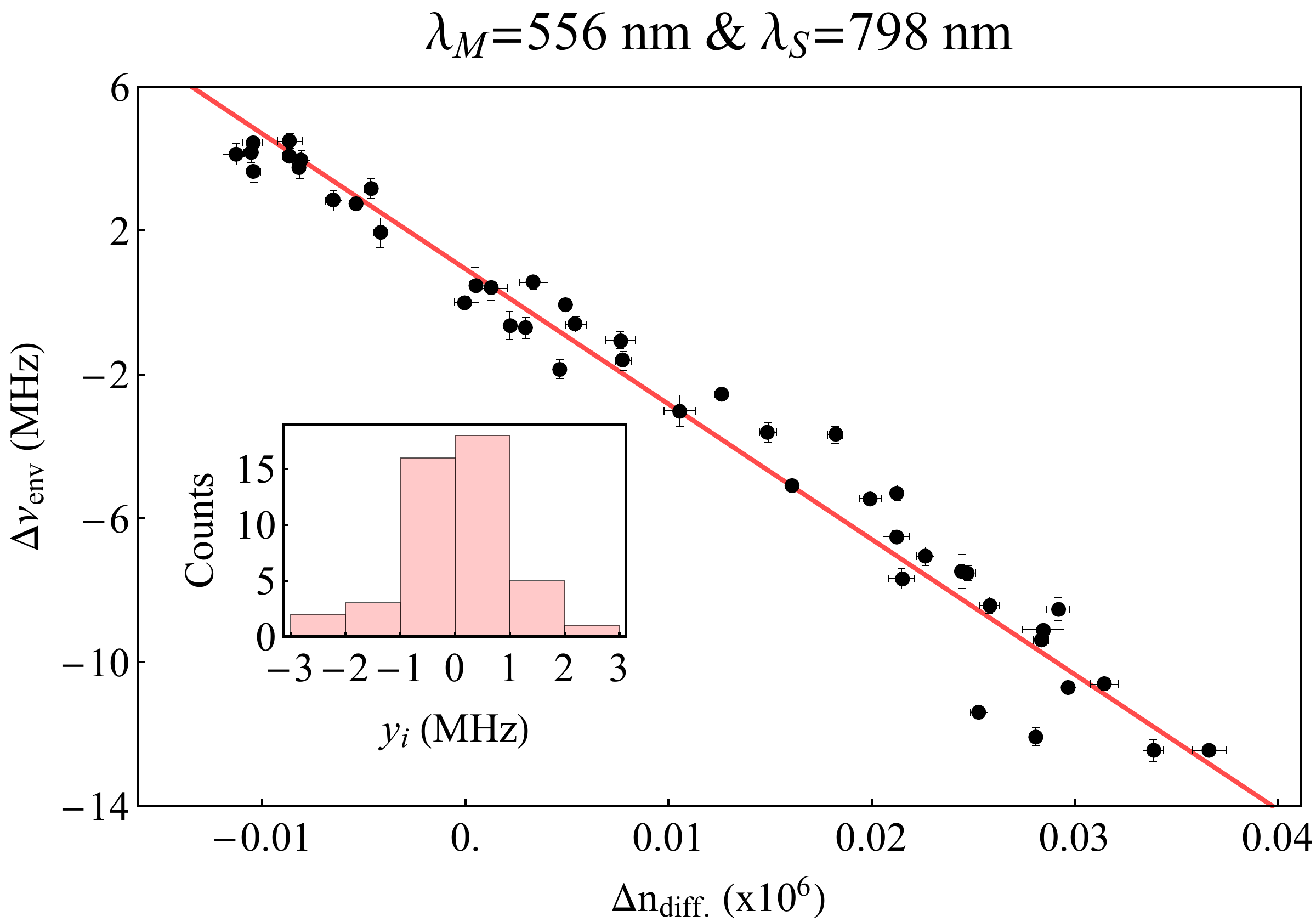}
		\centering
		\label{TPH}
		
	\end{subfigure}%
	~
	\begin{subfigure}[t]{0.5\textwidth} 
		\subcaption{}
		\includegraphics[height=2.4in]{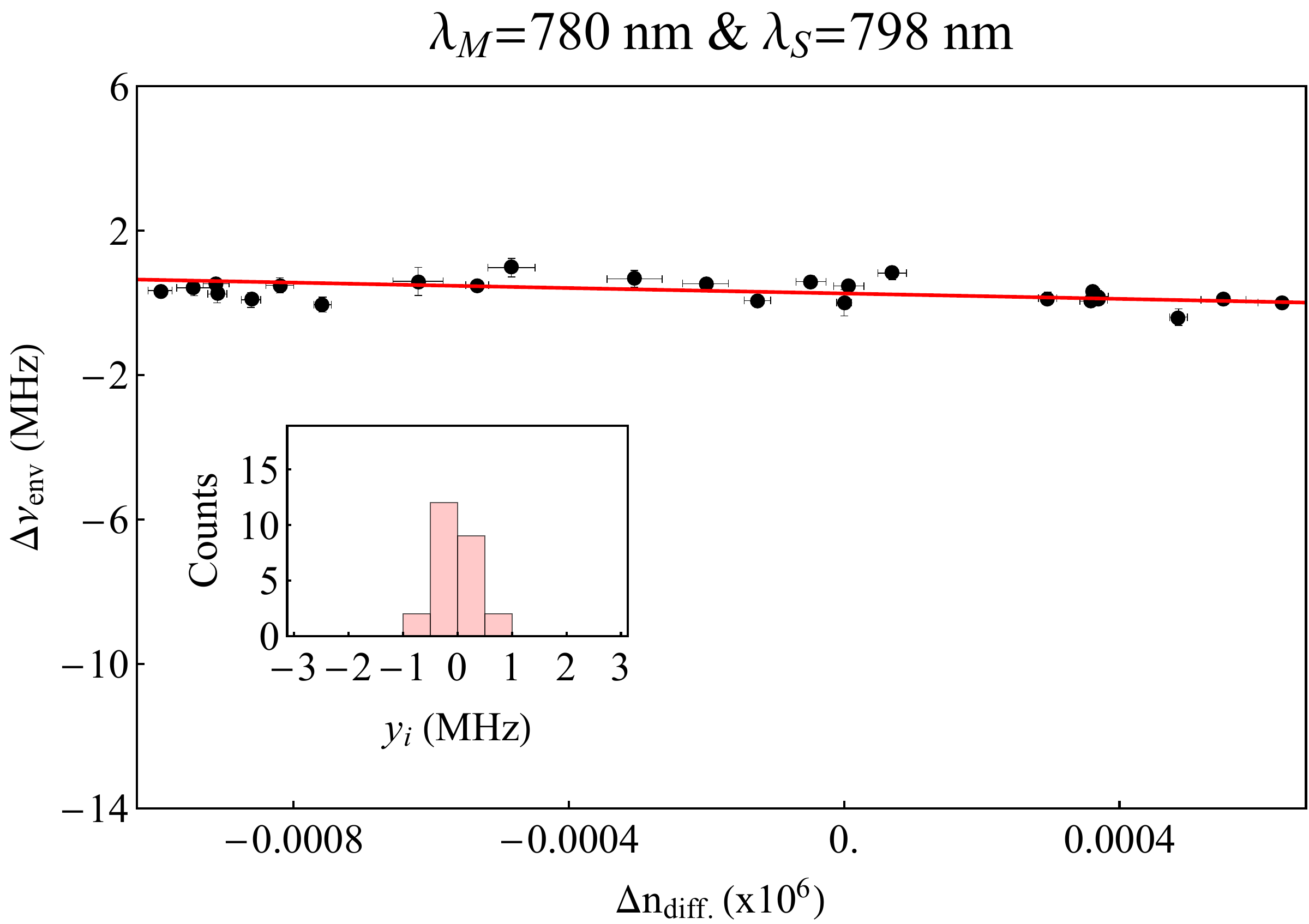}
		\centering
		\label{TPH780}
		
	\end{subfigure}

	\label{Env}
	\caption{Measurements of the shifts in the locked slave laser frequency, $\Delta\nu_{\textrm{env}}$, due to changes in the lab environmental conditions: 
		\textbf{(a)} Measurement of the change in setpoint $r$ such that the doubled slave laser is in resonance with the absolute frequency reference (the atomic line center of the $^1S_0|F=1/2,m_F=1/2\rangle\leftrightarrow {}^1P_1|F=3/2,m_F=3/2\rangle$ transition in $^{171}$Yb), as a function of change in differential refractive index, $\Delta n_{\textrm{diff.}}$, determined from measurements of pressure, temperature and humidity. In this measurement, we use $\lambda_M=556$~nm. The solid line has a slope given by the known value $ \nu^0_S=751.5/2$~THz. (The offset of the line was chosen to minimize the mean deviation of the measured points from the predicted line.) 
		Error bars along the $y$ axis are given by the fit error in the line center from Fig.~\ref{fig:scan} and along the $x$ axis by the propagated uncertainty in $\Delta n_{\textrm{diff.}}$ from random fluctuations in measurements from the BME280 sensor breakout board. \textbf{Inset}: Histogram of the deviations of the slave laser frequency, $y_i$, from the prediction for the $\lambda_M=556$ nm measurements. The deviations are normally distributed with 65\% probability, and with standard deviation of 1.0~MHz.
		\textbf{(b)} The same measurement as in Fig.~\ref{TPH}, but with $\lambda_M=780$~nm. The range of environmental conditions in this plot correspond to a drift of 13 MHz in Fig.~\ref{TPH}. \textbf{Inset}: For the $\lambda_M=780$~nm measurements, the deviations, $y_i$, are normally distributed with 69\% probability, and with standard deviation of 0.3~MHz.
	}
\end{figure*}

\subsection{Effect of the environment on the accuracy of the slave laser}\label{envt}
The expression for the slave laser setpoint $r$ (Eq.~\ref{ratio}) can be simplified to
\begin{align}
	r&=N_M-N_S\frac{ n_M \nu_M}{n_S\nu_S}\nonumber\\
	&\simeq N_M-N_S\frac{\nu_M}{\nu_S}[1+n_M-n_S]
\end{align}
since $n_M, n_S\simeq 1$.
Differentiating both sides of the equation yields,
\begin{equation}
\Delta r=N_S\frac{\nu_M}{\nu_S}\bigg(\frac{1+n_M-n_S}{\nu_S}\Delta\nu_S-\Delta(n_M-n_S)\bigg).
\label{driftdifferential}
\end{equation}
When the feedback loop is engaged ($\Delta r=0$), accuracy of the slave laser frequency ($\Delta\nu_S=0$) is only guaranteed when $\Delta(n_M-n_S)=0$.  The magnitudes of $n_i$ depend on environmental factors like temperature (T), pressure (P), humidity (H), and CO$_2$ content of air. Analytic expressions for this dependence of $n_i$ of air on T, P, H and CO$_2$ content is presented in Refs.~\cite{Edlen1966,Ciddor1996,Bonsch2003,Birch1993}. Eq.~\ref{driftdifferential} indicates that implementations of STCL in a cavity exposed to ambient air, long-term laser frequency stability and accuracy ($\Delta\nu_S=0$) requires that the setpoint is dynamically changed via feed-forward to account for variations in the lab environment i.e.
\begin{equation}
\Delta r=-\frac{N_S\nu_M}{\nu_S}\Delta(n_M-n_S).
\label{drifteqn}
\end{equation}
 Feed-forward is ideal for this application since changes in the ambient environmental parameters occur on a timescale of a few minutes, which is much slower than the bandwidth of the lock (10 ms). Along the lines of the work presented in Refs.\cite{Matsubara2005,Uetake2009}, we investigate  the effect of environmental parameters T, P, and H  on the slave laser frequency. The sensitivity of the slave laser frequency's dependence on T, P, or H , increases with increasing dissimilarity between the master and slave laser wavelengths.  
 
 We use the line center of the $^1S_0|F=1/2,m_F=1/2\rangle\leftrightarrow {}^1P_1|F=3/2,m_F=3/2\rangle$ transition in $^{171}$Yb as an absolute frequency reference (Fig.~\ref{fig:scan}) to determine the value of $\nu_S=\nu^0_S$, where $\nu^0_S=751 527 368.68(39)/2$~MHz~\cite{Kleinert2016} is half the reference transition frequency, since we frequency-double our slave laser for the atomic spectroscopy. We measure the effect of the environment (Eq.~\ref{drifteqn}) by experimentally determining the $r$ that brings the doubled slave laser into resonance with the atomic transition. We quantify drifts in the required $r$ by comparing it to an arbitrary reference $r_{\textrm{ref}}$:
\begin{align}
&\Delta r=-\frac{N_S\nu_M}{\nu^0_{S}}\Delta(n_M-n_S)\nonumber\\
\implies& {\beta} (r-r_{\textrm{ref}})=-\nu^0_S[(n_M-n_S)-(n_M-n_S)_{\textrm{ref}}],
\label{Edlen}\\
\implies& \Delta\nu_\textrm{env}=-\nu^0_S\Delta n_\textrm{diff.}.
\label{Edlen2}
\end{align}
where $r_{\textrm{ref}}=N_M- N_S\nu_M/\nu^0_S[1+(n_M-n_S)_{\textrm{ref}}]$ serves as a reference position of the atomic line center under the environmental conditions on an arbitrarily chosen day. 
 In this paper, we have used the method according to Ciddor\cite{Ciddor1996} (applicable over a wavelength range of 230 nm to 1690 nm) to perform all calculations related to the differential refractive index.   

The sensitivity of $\Delta\nu_{\textrm{env}}$ depends strongly on the difference between $\lambda_{M}$ and $\lambda_S$. From Eqs.~\ref{Edlen} and \ref{Edlen2}, the expected environmentally induced change $\Delta\nu_{\textrm{env}}$ is given by
\begin{align}
	\nonumber
	\Delta\nu_{\textrm{env}}  &=-\nu^0_S\Delta(n_M-n_S)\\ 
	&=\mathlarger{\mathlarger{\sum}}_{i}-\nu^0_S\frac{\partial (n_M-n_S)}{\partial X_i}\Delta X_i
	\nonumber
	\label{Edlendeif}
\end{align}
where $X_i$ is T, P, or H and $-\nu^0_S\partial (n_M-n_S)/\partial X_i$ is the sensitivity coefficient~\cite{Hamby1994} for the parameter $X_i$. For a given change, $\Delta X_i$ ,  
\begin{equation}
\Delta \nu_{X_i}=-\nu^0_S\frac{\partial (n_M-n_S)}{\partial X_i}\Delta X_i,
\end{equation}
with the partial derivatives evaluated at $\textrm{T} =22~^\circ$C, $\textrm{P}=101168$ Pa, $\textrm{H}=43.3$\%,  which are the mean values of the environmental parameters we explore in our measurements. In Fig.~{\color{magenta}4}, we plot the frequency shift $|\Delta \nu_{X_i}|$~\cite{Ciddor1996} resulting from a specific change of the parameter $X_i$ as a function of $\lambda_M/\lambda_S$. The values of $|\Delta \nu_{X_i}|$ decreases as $\lambda_M/\lambda_S$ approaches 1, implying that the closer the wavelengths $\lambda_M$ and $\lambda_S$ are to each other, the less sensitive is the accuracy of the slave laser frequency to variations in environmental parameters.

 The local lab environment is monitored using a BME280 sensor breakout board placed near the cavity. The sensor board measures the temperature, pressure and humidity of air which we average over the 8 minutes it takes to acquire a complete spectrum measurement. Spectrum measurements were taken over the course of a few weeks, during which the lab experienced a range of ambient environmental conditions. The largest contribution to changes in differential refractive index came from weather related atmospheric pressure changes ranging from 100323~Pa to 102224~Pa. Figs.~\ref{TPH} and ~\ref{TPH780} show the measured change in the lock point $r$ (scaled in frequency units) as a function of the change in the differential refractive index, $\Delta n_{\textrm{diff.}}$, calculated using the method according to Ciddor\cite{Ciddor1996} with the measured environmental conditions. The solid line in both plots has a slope given by the known resonance frequency $ \nu^0_S=751.5/2$~THz, and is offset vertically in each case to minimize the mean deviation of the points from the line predicted from the environmental conditions. Histograms of the deviations, $y_i$, from the theory are shown in insets of  Figs.~\ref{TPH} and ~\ref{TPH780}. The standard deviations are 1.0~MHz and 0.3~MHz, respectively, for the $\lambda_M=556$ nm and $\lambda_M=780$ nm data, suggesting that by using feed-forward based on the environmental measurements, drifts in the slave laser frequency (which typically occurs on the timescale of a few minutes) can be corrected in real time to that level of precision.

For $\lambda_M=780$~nm and $\lambda_S=798$~nm, the values of $|\Delta X_i|$ that induces a $\Delta\nu_{\textrm{env}}=1$ MHz are  $\Delta\textrm{T} =7.4~^\circ$C , $\Delta\textrm{P}=2390$~Pa, or $\Delta\textrm{H}=311$~\%. Such changes in temperature, pressure or humidity are never observed during the course of any one measurement shown in Fig.~\ref{TPH780} suggesting that the 0.3~MHz scatter in the 780~nm data is due to the error inherent to our experimental measurement, e.g. the error in fitting  to the line center (that is typically 250 kHz for $ \nu^0_S$) is due to number fluctuations between successive experimental realizations.

For $\lambda_M=556$~nm and $\lambda_S=798$~nm, the values of $|\Delta X_i|$ that induces a $\Delta\nu_{\textrm{env}}=1$ MHz are  $\Delta\textrm{T} =0.3~^\circ$C, $\Delta\textrm{P}=104$ Pa, or $\Delta\textrm{H}=14.6$\%. We attribute the increased error of 1.0~MHz in the $\lambda_M=556$~nm measurements to the facts that we do not control the atmospheric pressure in our lab (which can drift during the 8 min spectrum measurement time), and that we do not possess the level of precision in our temperature control needed to correct for fluctuations less than $0.3~^\circ$C. In addition, the cavity is susceptible to air currents, and we do not measure the temperature, pressure, and humidity inside the cavity. Subtracting out the environment insensitive measurement error of 0.3~MHz in quadrature, we estimate our error due to uncontrolled environmental parameters to be 0.9~MHz. 

\section{Summary and Outlook}
We have implemented an all-digital $\mu$C-based STCL with environmental monitoring of pressure, temperature and humidity. The environmental measurements have the precision to compensate for environmental drifts, with appropriate feed-forward to the slave laser setpoint. We demonstrate the capability to compensate for environmentally induced frequency drifts at the 0.9~MHz level for master and slave laser wavelengths of 556~nm and 798~nm as an example. Integration of the environmental sensors into the cavity could improve this performance. Currently the bandwidth of the STCL is limited by the frequency of the cavity PZT scan (100 Hz). Future implementations of the STCL may include designing custom cavities with PZT scan speeds in the kHz range, thereby increasing the bandwidth.
\section*{Acknowledgments}
We like to thank Tsz-Chun Tsui for his help in data acquisition. This work is supported by NSF PFC at JQI (Grant No. PHY1430094) and ONR (Grant No. N000141712411).


\begin{thebibliography}{21}
\bibitem [{dis()}]{discliamer}%
		\BibitemOpen
		\href@noop {} {\bibinfo  {journal} {The identification of commercial products
				is for information only and does not imply recommendation or endorsement by
				the Joint Quantum Institute or the National Institute of Standards and
				Technology}\ }\BibitemShut {NoStop}%
\end{thebibliography}
	%
	
\end{document}